\documentclass{iopconfser}

\usepackage{amsmath}
\usepackage{aas_macros}
\usepackage{graphicx}
\usepackage[dvipsnames]{xcolor}

\usepackage[normalem]{ulem}

\newcounter{mnotecount}
\newcommand{\mnote}[1]
{\protect{\stepcounter{mnotecount}}$^{\mbox{\footnotesize $\bullet$\themnotecount}}$ 
\marginpar{
\raggedright\tiny
$\!\!\!\!\!\!\,\bullet$\themnotecount: #1} }

\begin{document}

\title{Improving the inference of the stellar quantities using the extended $I$-Love-$Q$-$\delta M$ relations}

\author{Eneko Aranguren$^{1}$, José A. Font$^{2,3}$, Nicolas Sanchis-Gual$^{2}$, Raül Vera$^{1}$}

\affil{$^1$ Department of Physics, University of the Basque Country UPV/EHU, Bilbao, Spain}
\affil{$^2$ Departament d'Astronomia i Astrofísica, Universitat de València, València, Spain}
\affil{$^3$ Observatori Astronòmic, Universitat de València, València, Spain}

\email{eneko.aranguren@ehu.eus}

\begin{abstract}
In relativistic Astrophysics
the $I$-Love-$Q$ relations
refer to 
approximately EoS-independent relations involving
the moment of inertia, Love number, and quadrupole moment through some quantities that are normalised by the 
mass $M_0$ of the background configuration of the perturbative scheme.
Since $M_0$ is not an observable quantity, this normalisation hinders the direct applicability of the relations. A common remedy 
assumes that $M_0$ coincides with the actual mass of the star $M_S$; however, this approximation is only adequate for very slow rotation
(when the dimensionless spin parameter is $\chi_S<0.1$).
The more accurate alternative approach, based on the $I$-Love-$Q$-$\delta M$ set of relations, circumvents this limitation by enabling the inference of $M_0$. 
Here we review both approaches and provide numerical comparisons. 
\end{abstract}

\section{Introduction}
Due to the complexity of solving the Einstein Field Equations for rotating compact bodies, which in general do not have exact and analytical solutions (except for infinitesimally thin disks of dust~\cite{neugebauer1993}), relativistic stellar configurations are usually studied using either full numerical integration (see e.g.~\cite{RNS} for the RNS code, or~\cite{LORENE1,LORENE2} for the LORENE code), or semi-analytical approaches, often driven using perturbation theory in the framework of the so-called Hartle–Thorne model~\cite{hartlethorne}. Within the latter approach, the general setup is split into an interior problem and a vacuum exterior, which are matched at the timelike boundary of the star imposing the identification of the first and second fundamental forms at either side at each order in the rotation parameter. The rigorous treatment provided in \cite{Reina2015} has led to the first results on the existence and uniqueness of stationary and axially symmetric configurations to second order in perturbation theory \cite{MRV1,MRV2}. 

Perturbation theory is usually exploited up to second order for isolated rotating stars
(see \cite{aranguren:2022,Reina:2015_homogeneous,berti_2005}),
and up to first order for nonrotating stars embedded in a tidal field
(see \cite{Aranguren:2023,Hinderer:2009,diedrichs_2023}),
which is known as the electric-type tidal problem. In either scenario for a given equation of state (EoS) and central pressure $P_c$, the Tolman-Oppenheimer-Volkoff equations
\cite{tolman,oppenheimer_volkoff}
yield the mass $M_0(P_c)$ and radius $R_0(P_c)$ of the static and spherically symmetric background configuration. For a star rotating with angular velocity $\Omega_S$, the first order problem provides the moment of inertia $I_S$, related to the angular momentum $J_S(P_c,\Omega_S)$ via $I_S(P_c) = J_S(P_c,\Omega_S)/\Omega_S$. The second order problem produces two quantities: the mass contribution $\delta M(P_c)$, so that the total mass of the star is given by
\cite{hartlethorne}
\begin{align}
    M_S(P_c,\Omega_S) &= M_0(P_c) + \Omega_S^2 \,\delta M(P_c), \label{eq:MS}
\end{align}
and the quadrupole moment $Q_S$,
which measures the deformation of the gravitational field at the exterior, and is
related to the
eccentricity
of the star~\cite{hartlethorne}.
Regarding the tidal problem, the first order yields the leading order Love number $k_2(P_c)$, which quantifies how the shape of a spherically symmetric compact star is affected by the presence of a companion star, and is related to the
dimensionless
tidal deformability via\footnote{We take $G = c = 1$ throughout.} $\lambda_S(P_c) = (2/3) (R_0/M_0)^5 k_2(P_c)$
(see~\cite{flanagan_hinderer,Hinderer_2008} for more context).

In~\cite{yagi_yunes_iloveq_erratum}, a set of approximately EoS-independent relations involving the moment of inertia $I_S(P_c)$, the Love number [via $\lambda_S(P_c)$], and the quadrupole moment $Q_S(P_c,\Omega_S)$, was found for cold neutron stars and quark stars. These relations depend on the dimensionless and rotation-independent quantities
\begin{align}\label{eq:reduced}
    \overline{I}:=\frac{I_S}{M_0^3},\quad
    \overline{Q}:=\frac{Q_SM_0}{\Omega_S^2 I_S^2} = \frac{Q_S}{\chi_S^2 M_0^3},
\end{align}
together with $\lambda_S$. Note that we have used the definition of the dimensionless spin parameter $\chi_S:=I_S \Omega_S/M_0^2$. The reduced quantities~\eqref{eq:reduced} depend on the Tolman-Oppenheimer-Volkoff mass $M_0$, which differs from the actual mass of the star via~\eqref{eq:MS}. The fact that $M_0$ is not an observable magnitude hinders the applicability of the relations, since one must obtain (or fix) the value of $M_0$ in order to translate from the reduced $\{\overline{I},\overline{Q}\}$ to the actual $\{I_S,Q_S\}$.

In this regard, the approach usually followed in the literature (see for example, ~\cite{yagi_kyutoku_2014}) \ ---\ which we call the standard approach\ ---\ consists in fixing $M_0$ to the actual mass of the star, i.e.
\begin{align}\label{eq:M0std}
    M_0^\textup{std} = M_S.
\end{align}
Reference~\cite{yagi:2013} describes this as a ``small caveat'', justified by the fact that stars observed so far rotate slowly ($\chi_S<0.1$). However, recent detections of gravitational wave signals have found milisecond pulsars rotating at spins of $\chi_S\sim0.4$ (see~\cite{Hessels}). As argued in~\cite{Aranguren:dM}, this highlights the need to reassess the implications of the approximation $M_0=M_S$.

The procedure of this approach works as follows.
First, the $I$-Love and $Q$-Love relations are used to infer $\overline{I}$ and $\overline{Q}$ from $\lambda_S$, and then, with the values of $M_S$ and $\chi_S$ at hand, $I_S$ and $Q_S$ are obtained from~\eqref{eq:reduced} via
\begin{align}\label{eq:ISQS_standard}
        I_S^\textup{std} = M_S^3\,\overline{I},\quad Q_S^\textup{std} = \chi_S^2\,M_S^3\,\overline{Q}.
\end{align}
The implementation of this approach requires that $\lambda_S$, $M_S$ and $\chi_S$ be known. However, these quantities can be inferred from gravitational-wave observations of binary systems, since some related effective parameters
get imprinted in the waveform during the inspiral~\cite{abbott:2017,Zhu:2018,GW170817:measurements, GW190425}.

We stress that the universal relations hold as long as the quantities involved are normalised by $M_0$ (not $M_S$). Consequently, the standard approach results in deviations from universality, as was claimed already in~\cite{Doneva:2013rha}. The approximation $M_0 = M_S$ is only reasonable when $\Omega_S$ (or $\chi_S$) is small [see Eq.~\eqref{eq:MS}], which explains why the $I$–Love–$Q$ relations are typically applied to slowly rotating stars --- those rotating well below the Keplerian limit, which marks the upper bound for the applicability of the perturbative scheme.

Even if the universality is broken when taking $M_0 = M_S$, one may conjecture that there might still be one universal relation for each fixed value of some rotation magnitude. In~\cite{Chakrabarti:2013tca, Pappas:2013naa}, this was indeed shown to be the case for the $I$-$Q$ relation --- the universality was recovered after expressing the relation in terms of some dimensionless frequency and spin. However, this $I$-$Q$ relation is not directly applicable in observational astrophysics, since neither $I$ nor $Q$ can be extracted from gravitational wave data. As argued before, one needs to start from $\lambda_S$, so rotation-dependent $I$-Love and $Q$-Love relations are required, but neither of them has been established yet.

In~\cite{Aranguren:dM} we introduced an approach that dodges the approximation~\eqref{eq:M0std} by enabling the inference of $M_0$ in the first place. This is known as the extended approach, as it is based on an extended set involving the quantity $\delta M$. We describe it in what follows.

\section{Extended $I$-Love-$Q$-$\delta M$ relations}\label{sec:extended}
In \cite{reina:dM, Aranguren:dM} a new set of universal relations involving the normalised mass contribution
\begin{align}\label{eq:reduced_dM}
    \overline{\delta M} := \frac{M_S-M_0}{\Omega_S^2M_0^3\overline{I}^2} = \frac{M_S-M_0}{\chi_S^2 M_0},
\end{align}
was found. Given the value of $\lambda_S$, the $\delta M$-Love relation is used to extract $\overline{\delta M}$, from which the unique value of $M_0$ (denoted $M_0^\textup{ext}$) is obtained through
\begin{align}\label{eq:M0ext}
    M_0^\textup{ext} = \frac{M_S}{\overline{\delta M}\chi_S^2 +1}.
\end{align}
Once the value of $M_0^\textup{ext}$ has been inferred, the $I$-Love and $Q$-Love relations are then employed (just as in the standard approach) to obtain $\overline{I}$ and $\overline{Q}$, and using the $M_0^\textup{ext}$ extracted in~\eqref{eq:M0ext}, the stellar quantities $I_S$ and $Q_S$ are computed as
\begin{align}\label{eq:ISQS_extended}
    I_S^\textup{ext} = \left(M_0^\textup{ext}\right)^3\,\overline{I},\quad Q_S^\textup{ext} = \chi_S^2 \left(M_0^\textup{ext}\right)^3\,\overline{Q}.
\end{align}

\section{Comparison: standard approach vs extended approach}\label{sec:comparison}
\begin{figure}[htb]
\centering
    \includegraphics[width=\textwidth]{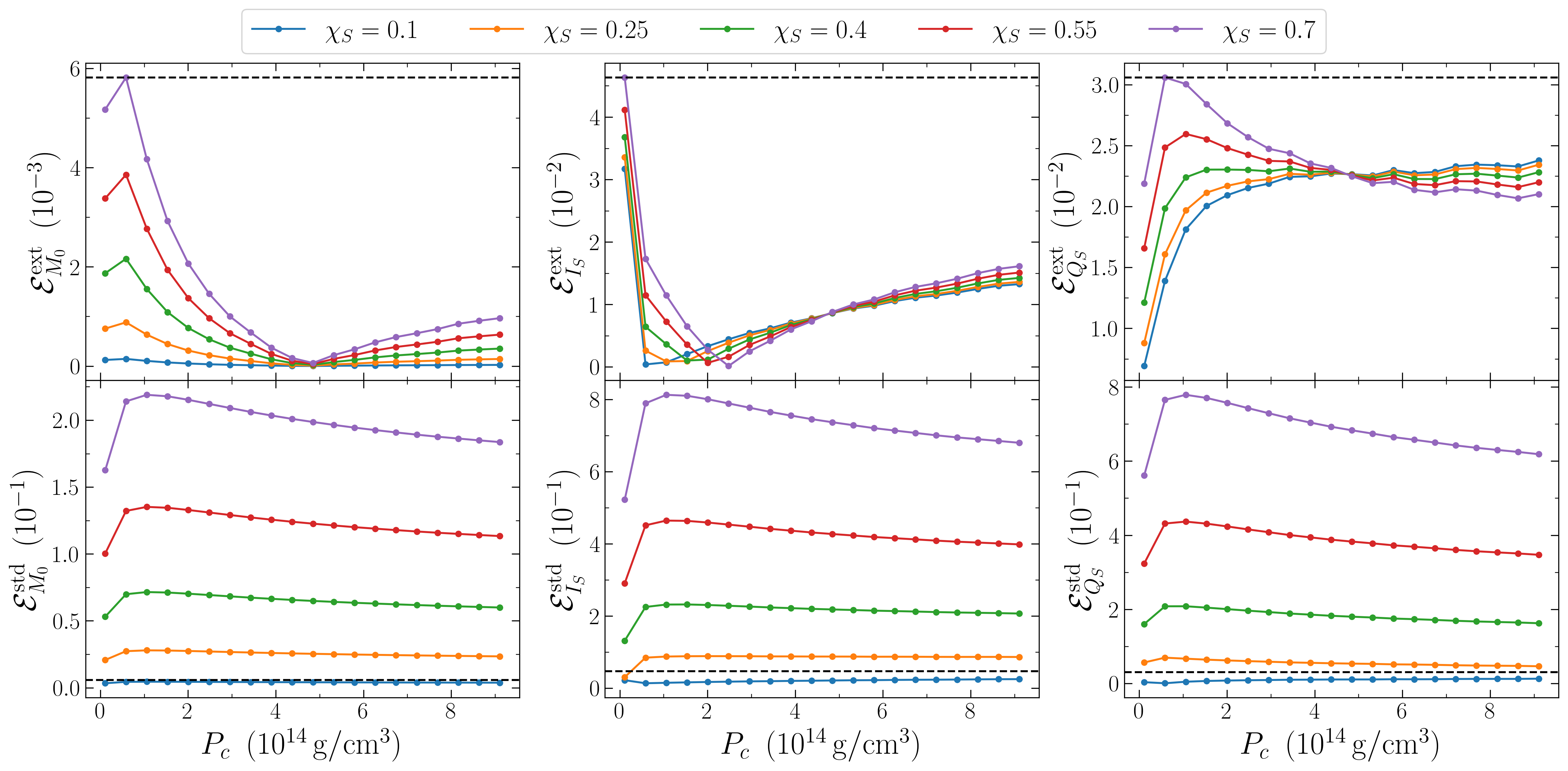}
    \caption{Relative errors of $M_0$ (left column), $I_S$ (middle column) and $Q_S$ (right column) as computed using perturbation theory, for the extended approach (upper row) and standard approach (lower row) as functions of $P_c$ and for different values of $\chi_S$. In order to compare the results from both approaches in a more visual way,
    a dashed line marks the highest error of the extended approach in every column.}
    \label{fig:relative_errors}
\end{figure}

Let us consider a polytropic EoS, where the energy density $\rho$ and pressure $P$ are related via $P = 100 \rho^2$.
We calculate the relative errors of $M_0$, $I_S$ and $Q_S$, defined as
\begin{align}\label{eq:relative_errors}
    \mathcal{E}_{M_0}^{\textup{std/ext}} = \frac{|M_0^{\textup{std/ext}} - M_0^\textup{H}|}{M_0^\textup{H}},\quad \mathcal{E}_{I_S}^{\textup{std/ext}} = \frac{|I_S^{\textup{std/ext}} - I_S^\textup{H}|}{I_S^\textup{H}},\quad \mathcal{E}_{Q_S}^{\textup{std/ext}} = \frac{|Q_S^{\textup{std/ext}} - Q_S^\textup{H}|}{Q_S^\textup{H}}
\end{align}
for both the standard (std) and the extended (ext) approaches, considering different values of the central pressure $P_c$ and spin parameter $\chi_S$. The magnitudes $M_0^\textup{H}$, $I_S^\textup{H}$ and $Q_S^\textup{H}$ correspond to the values of $M_0$, $I_S$, $Q_S$ as calculated using the
perturbative approach, respectively.
The results are shown in Fig.~\ref{fig:relative_errors}.

\section{Conclusions and comparison with previous results}

Figure~\ref{fig:relative_errors} shows that the extended approach yields more accurate results than its standard counterpart. The difference increases with $\chi_S$ ($\Omega_S$), as the approximation $M_0 = M_S$ becomes less accurate [see Eq.~\eqref{eq:MS}].
In addition, the relative errors of the extended approach exhibit a large variation, which may arise from the inherent noise of the universal relations themselves. In the standard approach, however, this effect appears to be overshadowed by the approximation $M_0 = M_S$, which introduces a larger source of error.

Let us note that the plots here differ from those in~\cite{Aranguren:dM} [see Fig.~1 therein], since the present calculations are based on $\chi_S$ rather than $\Omega_S$. To see the difference, compare~\eqref{eq:M0ext}, \eqref{eq:ISQS_extended}, \eqref{eq:ISQS_standard} here with (12), (13)-(14), (15)-(16) in~\cite{Aranguren:dM}, respectively.
Taking $\chi_S$ as the starting point, the value of $M_0^\textup{ext}$ depends solely on the $\delta M$-Love relation, while for the case of $\Omega_S$, it depends on both $\delta M$-Love and $I$-Love (via $\overline{I}$). Moreover, the quantity $\overline{I}$ also appears in $Q_S^\textup{std/ext}$ when using $\Omega_S$.

\section*{Acknowledgements}
We thank the anonymous referee for their useful comments and suggestions.
Work supported by the Spanish Agencia Estatal de Investigación (grants PID2021-125485NB-C21, PID2021-123226NB-I00 funded by MCIN/AEI/10.13039/501100011033 and ERDF A way of making Europe), by the Generalitat Valenciana (grant CIPROM/2022/49), by the Basque Government (IT1628-22), and by the European Horizon Europe staff exchange (SE) programme HORIZON-MSCA-2021- SE-01 (NewFunFiCo-101086251). EA is supported by the Basque Government Grant No. PRE\_2024\_2\_0078. NSG acknowledges support from the Spanish Ministry of Science and Innovation via the Ramón y Cajal programme (grant RYC2022-037424-I), funded by MCIN/AEI/10.13039/501100011033 and by “ESF Investing in your future”.

\bibliographystyle{iopart-num}
\bibliography{references}

\end{document}